\documentclass[prl,twocolumn,aps,superscriptaddress,notitlepage,floatfix,10pt]{revtex4-2}
\usepackage{amsmath,mathtools,amsthm,amssymb,pifont}
\usepackage[percent]{overpic}
\usepackage[utf8]{inputenc}
\usepackage{comment}
\usepackage[american]{babel}
\usepackage{graphicx,xcolor,bbold,titlesec}

\usepackage{MnSymbol}
\newtheorem{theorem}{Theorem}

\usepackage[colorlinks,
bookmarksopen,
bookmarksnumbered,
citecolor=teal,
linkcolor=teal,
urlcolor=teal]{hyperref}
\usepackage{braket}
\usepackage{enumitem}
\usepackage{subfigure}
\usepackage{ifthen}

\usepackage{orcidlink}
\usepackage{tikz,ifthen}
\usepackage{bbold}
\usepackage{tikz-network}
\usetikzlibrary{patterns,decorations.pathreplacing,calligraphy}
\usetikzlibrary{shapes,arrows.meta,decorations.pathmorphing}

\newcommand{\Tr}{\mathrm{Tr}}

\newcommand{\titleinfo}{Equivalence of quantum resources under ergodic dynamics}

\begin{document}
\title{\titleinfo} 

\author{Sreemayee Aditya~\orcidlink{0000-0002-0412-7944}}
\email{asreemay@uni-koeln.de}
\affiliation{Institut für Theoretische Physik, Zülpicherstraße 77a, 50937, Köln, Germany}

\author{Xhek Turkeshi~\orcidlink{0000-0003-1093-3771}}
\email{xturkesh@uni-koeln.de}
\affiliation{Institut für Theoretische Physik, Zülpicherstraße 77a, 50937, Köln, Germany}

\author{Piotr Sierant~\orcidlink{0000-0001-9219-7274}}
\email{piotr.sierant@bsc.es}
\affiliation{Barcelona Supercomputing Center Plaça Eusebi Güell, 1-3 08034, Barcelona, Spain}

\begin{abstract}
Quantum resource theories characterize distinct forms of nonclassicality in many-body quantum states, raising the question of whether these resources evolve independently under generic ergodic dynamics. 
Considering diagnostics quadratic in the state, we show that the dynamics of different resource measures become mutually interdependent and are governed by a few common degrees of freedom. 
For Haar-random circuits, the purity together with a single resource witness suffices to reconstruct the remaining resource measures, as we demonstrate for coherence, various asymmetries, and number entropies. The same relations hold, to a good approximation, under chaotic Floquet and continuous-time Hamiltonian dynamics, establishing that this dynamical resource equivalence extends beyond random circuits.
We further derive an exact coherence--imaginarity relation, remarkably accurate also beyond Haar-random circuits. 
Our results reveal an emergent simplification of many-body dynamics, in which sufficiently strong scrambling reduces seemingly distinct quantum resources to a few common dynamical degrees of freedom.
\end{abstract}

\maketitle
\paragraph{Introduction.}
Comprehending the non-classicality of many-body quantum states demands characterizing the correlation structures generated during their evolution. Anticoncentration~\cite{Dalzellanticoncentration2022,TurkeshiHilbert2024,ClaeysFock2025,Magni2025AnticoncentrationClifford, SauliereUniversality2025} and the emergence of randomness~\cite{BoulandOn2018,LamiAnticoncentration2025,FavaDesigns2025,ChengPseudoentanglement2025,IppolitiDynamical2023,ClaeysEmergentquantum2022,SchusterRandom2025,trigueros2026unitarydesignsdopedmatchgate} furnish prominent manifestations of the complexity of quantum states. 
Quantum resource theories~\cite{Chitambar2019} organize different forms of non-classicality, including entanglement~\cite{HorodeckiEntanglement2009,AmicoEntanglemennt2008}, nonstabilizerness~\cite{Veitch14,LiuWintermagic2022,leone2022stabilizer,Bravyi2016magic,Wang2019magic,Howard17rom,turkeshi2025magic}, coherence~\cite{Baumgratzcoherence2014,Streltsovcoherence2017,Saxenacoherence2020}, asymmetry~\cite{ares2023entanglement,aresasymmetry25,Aresasymmetryblackhole24,gotta2026enhancingentanglementasymmetryfragmented,aditya2026higherordersymmetricquantummpemba,turkeshi-24,summer2025resourcetheoreticalunificationmpemba,rac-25,castroalvaredo2026entanglementasymmetryrandomquantum}, imaginarity~\cite{imag1,imag2,imag3,imag4,imag5} and bosonic~\cite{ZhuangNG2018,TakagiNG2018} or fermionic~\cite{Hebenstreit2019all,Lumia24gaussian,Lyu24fermi,Coffman25fermi, sierantfermionicmagicresourcesquantum2025,haug2026practicaltestswitnessesfermionic,sierant2026theorymatchgatecommutant,ares2026asymmetrylowerboundfermionic,ares2026nongaussianityrandomquantumstates} non-Gaussianity according to the operational tasks they enable in several fields from quantum computation to metrology and sensing. 
Evolution of quantum resource measures in interacting many-body systems, therefore, offers a natural probe of how quantumness is dynamically generated, redistributed, and ultimately scrambled, connecting resource dynamics to quantum chaos~\cite{Haakebook,Lauchli08,kim13ballisticent,Leone21chaos,Varikuti25chaos}, information scrambling~\cite{Garcia23,Swingle16,Yunger19,Vikram24, Kaneyasu25,Odavic25},  thermalization~\cite{dalessio2016from,Pappalardi2022ETHFreeProbability,Abanin2019MBLColloquium,Sierant2025MBLReview,tirrito2025universalspreadingnonstabilizernessquantum,Jasser25,Bera25syk,Sticlet2025nonstabilizernesstransport,Falcao25mbl,Varikuti25deep}, critical phenomena~\cite{haug2023quantifying,tarabunga-24,TarabungaCastelnovo2024magicRokhsar,Collura2024NonStabFermGauss,Santra2025Latticegauge}, conformal field theory~\cite{White2021CFTemagical,Frau2025StabDisentanglingCFT,Hoshino25sre}, monitored dynamics~\cite{bejan2025magicspreadingunitaryclifford,Fux2024magictransition,Tirrito2025MagicGauss,Trigueros2025NoisyMagic,sierant2026theorymagicphasetransitions}, resource Mpemba effect~\cite{Aditya25Mbempa,xiao2026nonstabilizernessmpembaeffects} and metrological response~\cite{Hernandez25nonstab,lirasolanilla2026assemblingextensivequantumfisher}.

The dynamics of quantum resources are considerably harder to characterize than those of conventional observables, as resource monotones~\cite{Chitambar2019,CoeckeResourcetheory2016} are typically nonlinear in the evolving state. 
Random quantum circuits~\cite{Fisherrandomcircuits2023,Nahumoperatorspreading2018,Chanquantumchaos2018,Shivamquantumchaos2023} offer a natural framework for this problem, capturing ballistic entanglement growth using the minimal-membrane picture~\cite{Nahumentanglement2017,Zhou2019EmergentStatMech,Vasseur2019HolographicRTN,Sierant2023MinimalMembranes,Zhou2020EntanglementMembrane,Sommers2024ZeroTempMembranes} to information scrambling~\cite{Nahumoperatorspreading2018,vonKeyserlingk2018hydrodynamics}. Recent extensions to coherence~\cite{Bertoni2024ShallowShadows,TurkeshiHilbert2024,GarciaMartin2024SymplecticCircuits,Braccia2024ExactMoments,Christopoulos2025UniversalOverlaps,Magni2025AnticoncentrationClifford,ZhangEtAl2025DesignsMagicAugmented,Aditya26growthspreading}, non-Gaussianity~\cite{sierantfermionicmagicresourcesquantum2025,Aditya26growthspreading}, and nonstabilizerness~\cite{turkeshi2025magic,Aditya26growthspreading} have revealed a striking separation of timescales: while entanglement saturates over a time proportional to system size, certain resources do so logarithmically fast. Similar behavior occurs in ergodic Floquet systems, whereas Hamiltonian dynamics and conservation laws can lead to substantially slower resource generation~\cite{Tirritoanticoncentration2025,aditya2026coherencedynamicsquantummanybody}.

A recurring theme in nonequilibrium physics is that observables with disparate relaxation laws may nevertheless encode the same underlying dynamics: in classical glassy and aging systems, such observables synchronize when parametrized by an intrinsic clock rather than laboratory time, and entanglement has recently been shown to play this role in disordered systems~\cite{Evers_2023}. 
This raises a broader question: 
can apparently unrelated observables carry the same dynamical information? The question is particularly sharp for quantum resources, whose measures quantify distinct aspects of nonclassicality of a quantum state.

\begin{figure}[h]
\includegraphics[width=0.47\textwidth]{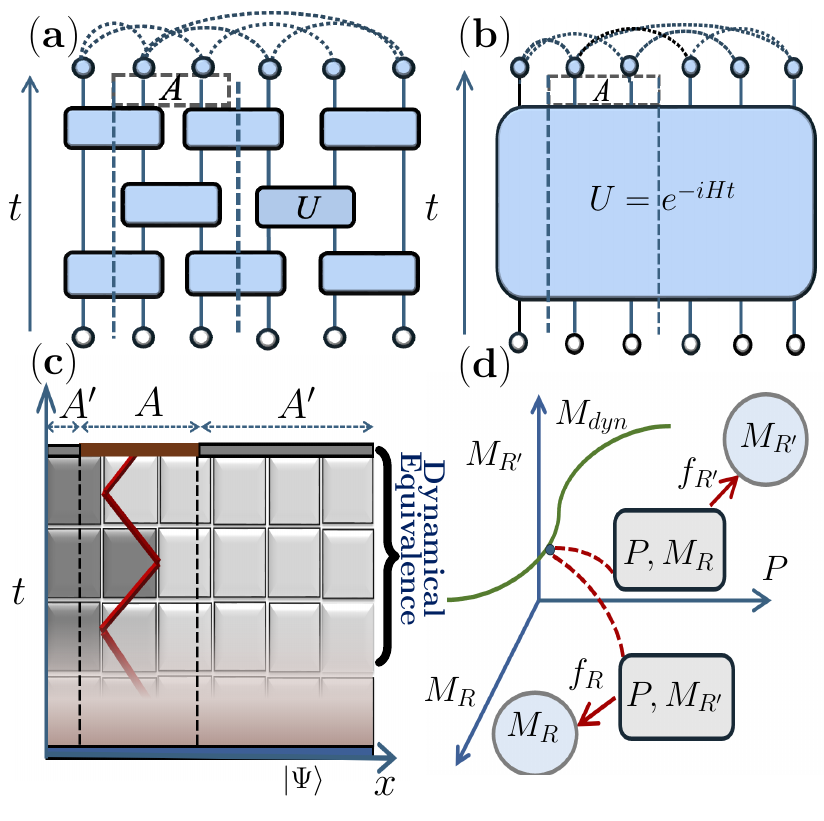}
\caption{\label{fig:schematic}
\textbf{Dynamical equivalence of resources.}
Ergodic dynamics generated by a local brick-wall circuit (a) or a global Hamiltonian dynamics (b) for a subsystem (A).
(c)~Resource measures that are quadratic in state correspond to distinct boundary states of the same replicated dynamics. To leading order, their evolution is governed by the uniform zero-wall configurations, fixed by the purity $P$, and a common one-wall mode (red trajectory) of the replica statistical mechanics.
(d)~The pair $({P,M_R})$ determines a second resource $M_{R'}$ through $f_{R'}$, while $({P,M_{R'}})$ determines $M_R$ through $f_R$, thus leading to dynamical equivalence between resources under generic ergodic evolution. $M_{\rm dyn}$: the resulting trajectory in the $(P,M_R,M_{R'})$ space.}
\end{figure}

In this Letter, we show that ergodic many-body dynamics generate equivalence relations between distinct resource measures. Focusing on
quantities that are quadratic in the state~\cite{turkeshi2026lecturenotesreplicatensor}, we organize
distinct diagnostics into a hierarchy in which the purity is the
least structured level and increasingly resource-sensitive functionals
resolve progressively finer information about the time-evolved state under generic ergodic dynamics.
Thus, quantities that are \textit{a priori} independent state functionals become
\emph{dynamically locked} under ergodic evolution. (Throughout, "equivalence" is meant in precisely this dynamical sense rather than as operational interconvertibility under tailored operations.) 
We demonstrate this equivalence across paradigmatic
ergodic settings---local Haar-random brickwork
circuits~\cite{Fisherrandomcircuits2023}, the Floquet kicked Ising
model~\cite{prosen98KIM,prosen99KIM,prosen2008kim}, and mixed-field
Ising model~\cite{kim13ballisticent}. We derive an exact
relation between coherence and imaginarity under Haar-random evolution
that remains accurate in other ergodic dynamics. These 
results therefore establish an emergent dynamical equivalence among inequivalent resource measures, see Fig.~\ref{fig:schematic}.

\paragraph{Quantities of interest.}
We consider a chain of $L$ qudits, with local Hilbert-space dimension
$q$, and a boundary subsystem $A$ comprising the first $L_A$ sites. Unless otherwise stated, we focus on the qubit case, $q=2$.
The subsystem Hilbert space has dimension $D=2^{L_A}$, and its reduced
density matrix at time $t$ is denoted by $\rho_A(t)$. We focus throughout on quantities that are quadratic in $\rho_{A}$. 
The coarsest diagnostic in this hierarchy is the purity
$P(t)={\rm Tr}(\rho_A(t)^2)$, or equivalently the R\'enyi-2 entropy, $S_2(\rho_A)=-\log_2 P$. (Logarithms are in base $q$ throughout.)
The finer diagnostics are drawn from quantum resource theories, which
quantify nonclassical features of a state by distinguishing a set of
\emph{free states}, such as those easy to implement or prepare, and \emph{free operations},
which cannot generate the resource~\cite{Chitambar2019}. Any state
outside the free set is defined as resourceful. Good measures of resources, the so-called resource monotones, vanish on
free states and are nonincreasing under free operations. Following
Ref.~\cite{Aditya26growthspreading}, we quantify each resource through
the R\'enyi-2 relative-entropy
\begin{equation}
M_2(\rho_A)=S_2\!\left(\rho_A^{F}\right)-S_2(\rho_A),
\label{eq:monotone}
\end{equation}
where $\rho_A^{F}$ is the free reference state obtained by projecting
$\rho_A$ onto the corresponding free set. Since these monotones are
normalized by the purity, we call them \emph{purity-normalized} quantities. For
coherence, the free set is the convex hull of the computational basis
projectors,
${\rm INCOH}\equiv\{\sum_{z}p_z\ket{z}\!\bra{z}:\,p_z\geq0,\,\sum_zp_z=1\}$~\cite{Baumgratzcoherence2014,Streltsovcoherence2017,Saxenacoherence2020,Levi14};
the free reference is the dephased state
$\rho_{A}^{D}=\sum_{z}\langle z|\rho_{A}|z\rangle |z\rangle \langle z |$,
and Eq.~\eqref{eq:monotone} yields the R\'enyi-2 relative entropy of
coherence
\begin{equation}
C_d
=
S_2\!\left(\rho_{A}^{D}\right)-S_2(\rho_A).
\label{eq:Cd}
\end{equation}
For imaginarity, the free set consists of the states that are real in the
computational basis,
${\rm REAL}\equiv\{\rho:\,\langle z|\rho|z'\rangle\in\mathbb R\ \ \forall
z,z'\}$~\cite{imag1,imag2,imag3,imag4}; the free reference is the
realification $\mathfrak{R}(\rho)=(\rho+\rho^{T})/2$, with transposition
taken in the computational basis, giving the R\'enyi-2 relative entropy
of imaginarity
\begin{equation}
I(\rho_A)
=S_2(\mathfrak{R}(\rho_A))-S_2(\rho_A).
\label{eq:Ci}
\end{equation}
For asymmetry under a symmetry group $G$, the free set consists of the
$G$-invariant states,
${\rm SYM}_G\equiv\{\rho:\,U_g\rho U_g^\dagger=\rho\ \ \forall g\in
G\}$~\cite{ares2023entanglement,aresasymmetry25,Aresasymmetryblackhole24,gotta2026enhancingentanglementasymmetryfragmented,aditya2026higherordersymmetricquantummpemba,turkeshi-24,summer2025resourcetheoreticalunificationmpemba};
the free reference is the twirled state $\rho_{A,G}$, giving the
R\'enyi-2 entanglement asymmetry
\begin{equation}
\Delta S_G
=
S_2(\rho_{A,G})-S_2(\rho_A).
\label{eq:asym}
\end{equation}
We consider various asymmetries: for the $U(1)$ charge $Q_A=\sum_{j\in A}\hat n_j$, the twirled state is $\rho_{A,Q}
=
\sum_Q\Pi_Q\rho_A\Pi_Q$,
where $\Pi_Q$ projects onto the charge-$Q$ sector, yielding the charge
asymmetry $\Delta S_{Q}$~\cite{ares2023entanglement,aresasymmetry25,Aresasymmetryblackhole24,summer2025resourcetheoreticalunificationmpemba,turkeshi-24}. 
We also introduce the subsystem dipole
moment $P_A=\sum_{j=1}^{L_A}j\,\hat n_j$
and define the dipole asymmetry $\Delta S_{P}$~\cite{gotta2026enhancingentanglementasymmetryfragmented,aditya2026higherordersymmetricquantummpemba} by twirling
$\rho_A$ over the eigenspaces of $P_A$. 

Besides the purity-normalized quantities, we consider diagnostics of the charge and
dipole distributions that are not normalized by $P$. Let $p_Q={\rm Tr}(\Pi_Q\rho_A)$
be the probability of measuring charge $Q$ in $A$, then the R\'enyi-2 number entropy and subsystem charge fluctuation can be defined as 
\begin{equation}
S_{Q}
=
-\log_q\sum_Qp_Q^2,
\qquad
\operatorname{Var}Q_A
=
\langle Q_A^2\rangle-\langle Q_A\rangle^2.
\label{eq:num}
\end{equation}
Similarly, one can define analogous quantities for the dipole case.
All these resource-sensitive diagnostics are therefore functionals of the same state.

\emph{Setups.---} To examine the dynamical equivalence of resources, we consider three paradigmatic setups of ergodic many-body dynamics. As the minimal setup, we  
consider a one-dimensional
chain of $L$ qubits, evolving under brick-wall
circuit~\cite{Fisherrandomcircuits2023}. The evolution operator thus reads $U=\prod_{r=1}^{t}U_{r}$
,
where $t$ is the circuit depth, referred to as time. The
layers $U(r)$
of the circuits are fixed as
\begin{equation}
U(2m)
=
\prod_{i}^{N/2-1}
u_{2i,2i+1} , \qquad U(2m+1)=
\prod_{i=1}^{N/2}
u_{2i-1,2i}
\end{equation}
where each two-qubit gate $u_{i,j}$ is drawn from the unitary Haar random ensemble $U(q^2)$, and the initial state is chosen to be the product state $\ket{0\cdots 0}$. 
As a second setup, we consider the \emph{kicked Ising model} (KIM)~\cite{prosen2008kim,prosen98KIM,prosen99KIM}
with Floquet operator
\begin{equation}
U_{\mathrm{KIM}} \;=\;
e^{-i b \sum_j X_j}\,
e^{-i \left( \sum_j h_j Z_j + J \sum_j Z_j Z_{j+1} \right)},
\label{eq:KIM}
\end{equation}
at the quantum-chaotic parameter point $J = 1$, $b = (5+\sqrt{5})/8$, and
$h_j = (1+\sqrt{5})/4$, evolved from random product states
$\ket{\Psi_0} = \prod_{j=1}^{L} U^{(1)}_j \ket{0}^{\otimes L}$, with
$U^{(1)}_j$ independent Haar-random single-qubit unitaries.
The time evolved state $U_{\mathrm{KIM}} ^t\ket{\Psi_0}$ is obtained by employing the
fast Hadamard transformation~\cite{Lezama19fasthadamard,sierant23fasthadamard}. Finally, we consider continuous-time evolution under a variant of the
\emph{mixed-field Ising model} (MFIM)~\cite{kim13ballisticent}
\begin{eqnarray}
H&=& \sum_{j=1}^{L}(h+(-1)^j h_{st}) Z_j
+J\sum_{j=1}^{L-1} Z_jZ_{j+1}
+b\sum_{j=1}^{L} X_j\nonumber\\
&&+D_{M}\sum_{j=1}^{L-1}
\left(X_jY_{j+1}-Y_jX_{j+1}\right),
\end{eqnarray}
with \(J=1\), \(h=(1+\sqrt5)/4\), \(b=(5+\sqrt5)/8\), $h_{st}=0.3$, and
\(D_{M}=0.6\); the last term is a uniform Dzyaloshinskii--Moriya (DM)
interaction with DM vector along the spin-\(z\) direction. For
\(D_{M}=0\) the Hamiltonian is real in the \(Z\)-basis and exhibits
Gaussian-orthogonal-ensemble (GOE) spectral
statistics~\cite{Mehta1990,Haakebook} in the chaotic regime. The DM term
introduces imaginary matrix elements and, combined with the staggered
field at even \(L\), breaks all antiunitary symmetries, yielding
Gaussian-unitary-ensemble (GUE) level
statistics~\cite{Mehta1990,Haakebook} analogous to the unitary Haar
random case. We evolve $z$-basis initial states with energy in the middle
of the many-body spectrum,
$|\langle \Psi_0 | H | \Psi_0 \rangle - \bar{E}\,| / (E_{\max} - E_{\min})
\leq 0.05$ with $\bar{E} = (E_{\max} + E_{\min})/2$, using the Chebyshev
time-evolution method~\cite{TalEzer84, Weisse06, sierant22challengesMBL}, and adopt open
boundary conditions (OBCs) in all cases.

\paragraph{Resource equivalence from replica dynamics.}
We now explain how distinct quadratic resource measures become mutually
dependent under Haar-random evolution. Let $R$ label a resource and $M_R$ its measure, with
$M_R\in\{C_d,\Delta S_Q,\Delta S_P,S_Q\}$. Since the circuit gates and/or initial states are sampled randomly,
nonlinear resources admit inequivalent ensemble averages. We therefore define the
quenched and annealed measures
\begin{align}
M_R^{\rm que}
&=-\mathbb E_{\Omega}\log_q
\frac{\mathcal Z_R^{(\Omega)}(t)}{\eta_R^{(\Omega)}(t)},
~~
M_R^{\rm ann}
&=-\log_q
\frac{\mathbb E_{\Omega}\mathcal Z_R^{(\Omega)}(t)}
{\mathbb E_{\Omega}\eta_R^{(\Omega)}(t)}.
\label{eq:annealed_quenched}
\end{align}
The quenched average characterizes typical realizations, while the
annealed quantities is directly accessible within the replicated description.
Since the underlying second moments demonstrate a strong self-averaging effect in the chaotic regime,
$M_R^{\rm ann}$ furnishes an informative proxy for $M_R^{\rm que}$, with their
difference controlled by finite-size fluctuations. In addition, exact
state-vector simulations of the chaotic Floquet and Hamiltonian models showcase
that these annealed relations hold more generally.

Now although $M_R$ is nonlinear
in $\rho_A$, its annealed second moment is a linear function of
the two-copy averaged state $X_t=\mathbb E_U[\rho_A(t)^{\otimes2}]$. Denoting $\mathcal Z_R(t)={\rm Tr}(B_R X_t)$, we get 
$2^{-M_R(t)}=\frac{\mathcal Z_R(t)}{\eta_R(t)}$,
where $B_R$ is the boundary operator fixed by the resource measure analyzed. Here, $\eta_R=P$ for purity-normalized quantities, while $\eta_R=1$ for purity-unnormalized ones. Details of the construction
are given in the Supplemental Material~\cite{SM}.

A Haar random gate on two
neighboring sites projects~\cite{Nahumentanglement2017,Zhou2019EmergentStatMech} the 
two-copy (or replica) state $\rho_A^{\otimes 2}$ onto the
identity and swap configurations, \(e\) and \(s\), so the replicated
dynamics admit a domain-wall description: an \(e\)-\(s\) interface
propagates, annihilates with another wall, or is absorbed at an open
boundary, but is never created. The transfer matrix is therefore
triangular in domain-wall number, and its dynamics is
controlled, to leading order, by the zero- and one-wall sectors.

The zero-wall sector contains only the uniform configurations
$e_A=\bigotimes_{j\in A}e_j$ and $s_A=\bigotimes_{j\in A}s_j$:
$X_t^{(0)}=a_e e_A+a_s s_A$. The coefficients are fixed by
normalization and the measured purity \(P(t)={\rm Tr}(s_A X_t)\) as
$a_e=\frac{D-P}{D(D^2-1)}$, $a_s=\frac{DP-1}{D(D^2-1)}$. Hence, at zeroth order
every resource measure is a function of the purity alone. The boundary operators enter through the contractions $n_e^R={\rm Tr}(B_R e_A)$ and $n_s^R={\rm Tr}(B_R s_A)$; for all purity-normalized quantities considered here,$n_e^R=D$.
For the purity-normalized quantities we get
\begin{equation}
2^{-M_R ^{(0)}}
=
\frac{D^2-n_s ^R}{P\,D(D^2-1)} \,+\,\frac{n_s ^R-1}{D^2-1}.
\label{eq:purity_law_main}
\end{equation}
The purity-unnormalized quantities can be recast as
\begin{eqnarray}
2^{-S_{Q}^{(0)}}
&=&\frac{\sigma(D-P)+D(DP-1)}
        {D(D^2-1)},\nonumber\\
\langle\operatorname{Var}Q_A\rangle^{(0)}
&=&\frac{L_A D(D-P)}{4(D^2-1)},                        \label{eq:Var0}
\end{eqnarray}
with $\sigma=\binom{2L_A}{L_A}$.
Eqs.~\eqref{eq:purity_law_main}--\eqref{eq:Var0} constitute the
purity-controlled, zeroth-order description. 
In particular, the purity-normalized resources $\propto 1/P$, while the purity-unnormalized ones $\propto P$.

The leading correction to~\eqref{eq:purity_law_main}-\eqref{eq:Var0}  is the slowest one-wall mode, represented schematically in Fig.~\ref{fig:schematic}(c), upon whose inclusion
$X_t\simeq a_e e_A+a_s s_A+\mathcal A(t)\mathcal{Q}_p$, with $p=t\bmod2$, where
\(\mathcal{Q}_p\) is a fixed wall profile on the parity-selected bonds due to the brickwall evolution. The
lowest open-chain mode \(\phi_1(b)\propto\sin[\pi(b+1)/L]\) within the
bare one-wall sector and the amplitude \(\mathcal A(t)\) carry the
full time dependence. A single wall occurs in two orientations,
$s\cdots e$ and $e\cdots s$, the one whose environment lies in the swap
domain being suppressed by the bath factor $\kappa=q^{-(L-L_A)}$ upon
tracing out the complement. In addition, the uniform components of \(\mathcal{Q}_p\) are fixed
by ${\rm Tr}(\mathcal{Q}_p)={\rm Tr}(s_A \mathcal{Q}_p)=0$, so the one-wall correction
changes neither normalization nor the measured purity.

Therefore, contracting the fixed profile with
$B_R$ gives a time-independent sensitivity $\mathcal N_R$. To display
the origin of its terms explicitly, let $c_R(m)$ denote the contraction
with a one-wall configuration containing $m$ swap sites and define their positions
$\mathcal {M}_p=\{1\leq m\leq L_A-1:m\equiv p+1\ ({\rm mod}\ 2)\}$ and we also further define
$\mathfrak G_p(x)=\sum_{m\in\mathcal M_p}\sin(\pi m/L)x^m$. The 
coefficient can thus be written as
\begin{eqnarray}
\mathcal N_R
&=&{}
\underbrace{\mathcal F_R}_{\substack{\text{dominant wall}\\[-1pt]s\cdots e}}
+
\underbrace{2^{-(L-L_A)}\mathcal F_{\widetilde R}}_
{\substack{\text{bath-suppressed wall}\\[-1pt]e\cdots s}},\nonumber\\
\mathcal F_R
&=&{}
\underbrace{
\frac{1}{D}\sum_{m\in\mathcal M_p}
\sin\!\left(\frac{\pi m}{L}\right)c_R(m)
}_{\text{one-wall boundary contribution}}
+\underbrace{
\frac{\mathfrak G_p(2)\big(n_e^R-Dn_s^R\big)}{D(D^2-1)}
}_{\text{uniform $e_A$-sector correction}}
\nonumber\\
&&+
\underbrace{
\frac{\mathfrak G_p(1/2)\big(n_s^R-Dn_e^R\big)}{D^2-1}
}_{\text{uniform $s_A$-sector correction}}.
\label{eq:one-wall-sensitivity}
\end{eqnarray}
Here $\widetilde R$ denotes the transposed boundary readout computed by taking
$c_{\widetilde R}(m)=c_R(L_A-m)$ and
$(n_e^{\widetilde R},n_s^{\widetilde R})=(n_s^R,n_e^R)$. 
All time dependence is therefore carried by the single amplitude
$\mathcal A(t)$, which followed by first-order derivation factorizes as
\begin{equation}
\Delta_R(t)\equiv
2^{-M_R(t)}-2^{-M_R ^ {(0)}}
=\mathcal A(t)\frac{\mathcal N_R}{\eta_R(t)},
\label{eq:deviation_main}
\end{equation}
Eliminating the common amplitude with a witness $R'$ then yields
\begin{equation}
2^{-M_R(t)}\simeq 2^{-M_R^{(0)}}
+\rho_{RR'}\,\Delta_{R'}(t),
\label{eq:equivalence_main}
\end{equation}
with \(\rho_{RR'}(P)=[\eta_{R'}(P)/\eta_R(P)]
(\mathcal N_R/\mathcal N_{R'})\), for witnesses with $\mathcal N_{R'}\neq 0$.
The normalization ratio is unity within two purity-normalized or two purity-unnormalized quantities,  while an unnormalized target with a normalized witness acquires one factor of $P$ ($P^ {-1}$ with the
roles reversed).

\paragraph{Random circuit dynamics.} 
We now test the resource equivalence relations in Haar-random brick-wall circuits. 
Throughout, $f_R(P)$ denotes the purity-only
reconstruction of the resource $M_R$, obtained from
Eq.~\eqref{eq:purity_law_main} with the measured purity as the sole
input, whereas $f_R(P,R')$ denotes the first-order reconstruction of
Eq.~\eqref{eq:equivalence_main}, in which the measure $R'$ additionally
fixes the common one-wall amplitude. 
The main text focuses on the purity-normalized
pair $(C_d,\Delta S_Q)$.

\begin{figure}[htb]
\centering
\includegraphics[width=0.495\textwidth]{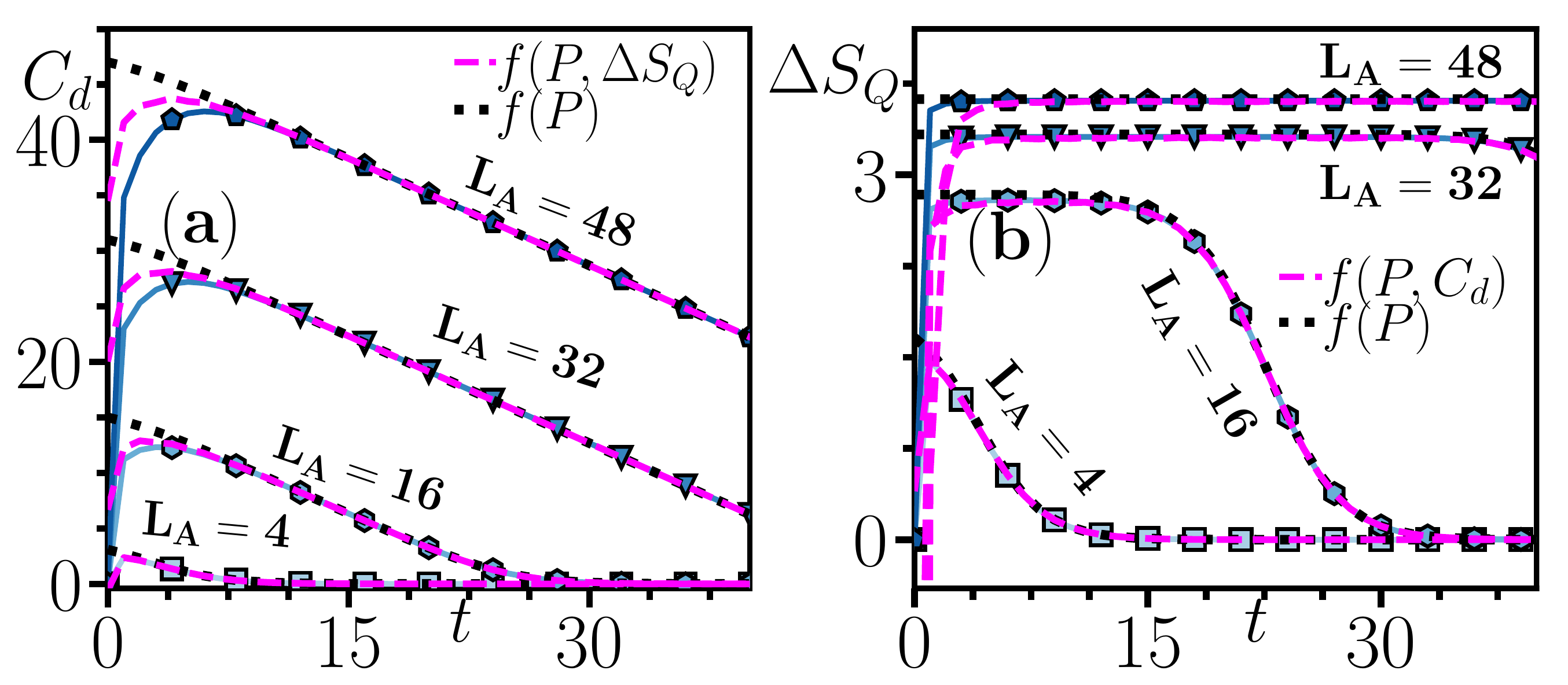}
\caption{\label{fig:rtn}
\textbf{Dynamical equivalence in Haar-random circuits.}
(a)~Annealed second-R\'enyi coherence $C_d$ reconstructed using the
charge asymmetry $\Delta S_Q$ as a witness; (b)~the
reconstruction of $\Delta S_Q$ using $C_d$. We consider a random brick-wall circuit with
$L=256$ and $L_A=4,16,32,48$. Markers denote numerical
data obtained within the tensor-network converged time window. Black dotted lines show the purity-only prediction $f_R(P)$; magenta dashed lines show
the first-order prediction $f_R(P,R')$ from
Eq.~\eqref{eq:equivalence_main}.}
\end{figure}

Figure~\ref{fig:rtn} demonstrates the 
replica tensor network~\cite{turkeshi2026lecturenotesreplicatensor} calculations for systems of $L=256$ qubits:
panel (a) utilizes $\Delta S_Q$ to predict $C_d$, while
panel (b) uses $C_d$ to predict $\Delta S_Q$. The purity-only curves
$f_R(P)$ capture the relaxation envelope but miss the initial rise
and the resource-dependent transient. Supplying the one-wall amplitude
through the other resource removes this discrepancy to some extent: the first-order
curves $f_R(P,R')$ follow the numerical data, reproducing the maximum and subsequent decay
of $C_d$ as well as the subsystem-dependent plateaus and relaxation of
$\Delta S_Q$. Thus, two inequivalent purity-normalized quantities 
become mutually predictable once the purity and one of the resource measures are known.

\paragraph{Ergodic many-body dynamics.}
Our resource equivalence relations~\eqref{eq:purity_law_main}, \eqref{eq:Var0}, \eqref{eq:equivalence_main} stem from the projection onto the span of identity and swap configurations which arises due to the Haar randomness of the circuit gates~\cite{turkeshi2026lecturenotesreplicatensor}. The dynamics of ergodic many-body systems is considerably more complex. Hence, the following analyses of the KIM and MFIM are more stringent tests of the robustness and universality of the uncovered relations. Importantly, the purity and the Haar-derived sensitivities $\mathcal N_R$ enter these comparisons as a parameter-free ansatz (no quantity is fitted); thus, the agreement below constitutes a genuine prediction of our dynamical equivalence result. 

\begin{figure}[htb]
\centering
\includegraphics[width=0.49\textwidth]{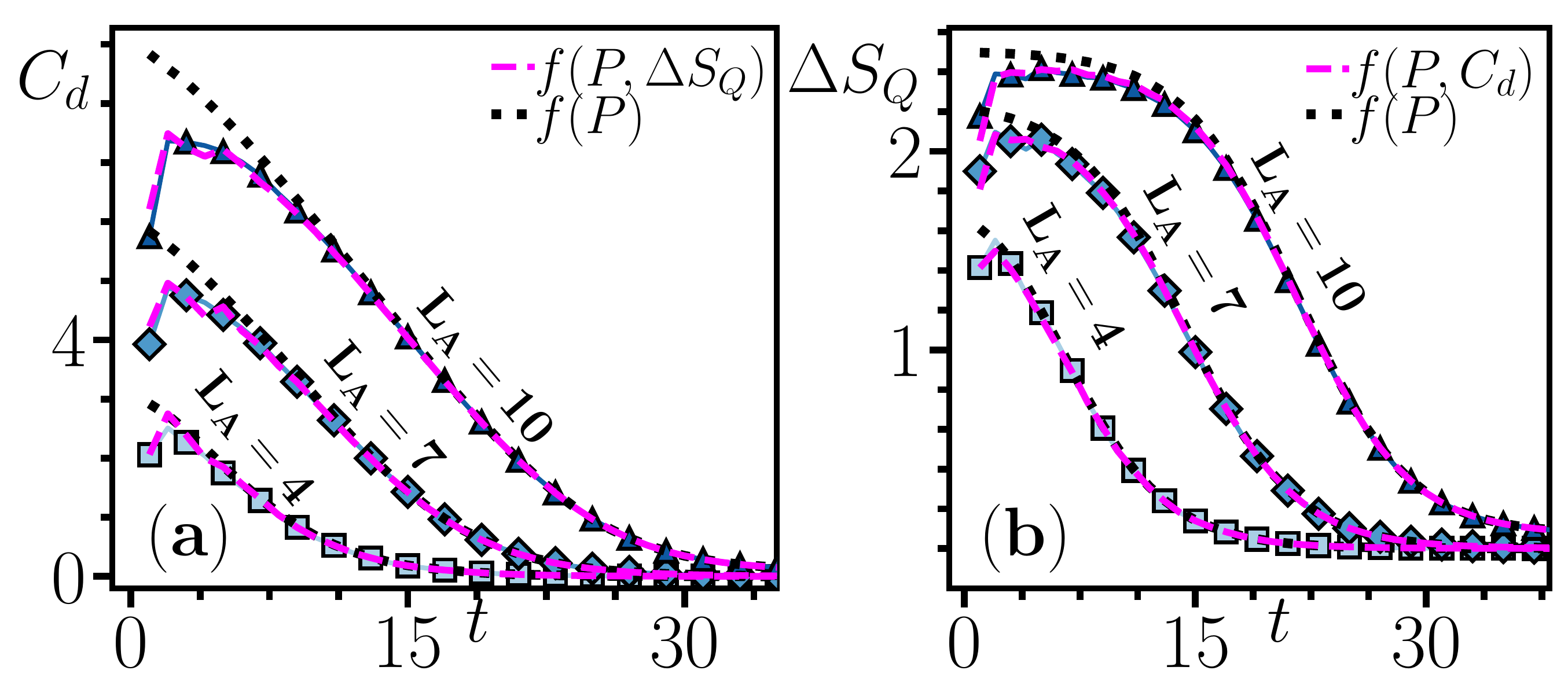}
\caption{
\textbf{Dynamical equivalence in the kicked Ising model.}
Same analysis as Fig. \ref{fig:rtn} in KIM (a) relative entropy of coherence $C_d$ reconstructed from $\Delta S_Q$;
(b)~charge asymmetry $\Delta S_Q$ reconstructed from $C_d$. In this case, we consider $L=24$ and $L_A=4,7,10$. 
Notations as in Fig. \ref{fig:rtn}.}
\label{fig:kim}
\end{figure}

Figure~\ref{fig:kim} presents the results for the chaotic
KIM, with the
underlying second moments averaged over $100$ random product states
before taking the logarithm. In both panels, the purity-only prediction
overestimates the early-time resource and approaches the numerical
relaxation only at later times. By contrast, the first-order
reconstruction from the complementary resource follows the full
transient: $\Delta S_Q$ reproduces the growth and decay of $C_d$ in
panel (a), while $C_d$ reconstructs the early-to-late time behavior of
$\Delta S_Q$ in panel (b). The agreement for all displayed subsystem
sizes shows that the two-resource reduction is not limited to the Haar random circuit dynamics.

\begin{figure}[htb]
\centering
\includegraphics[width=0.49\textwidth]{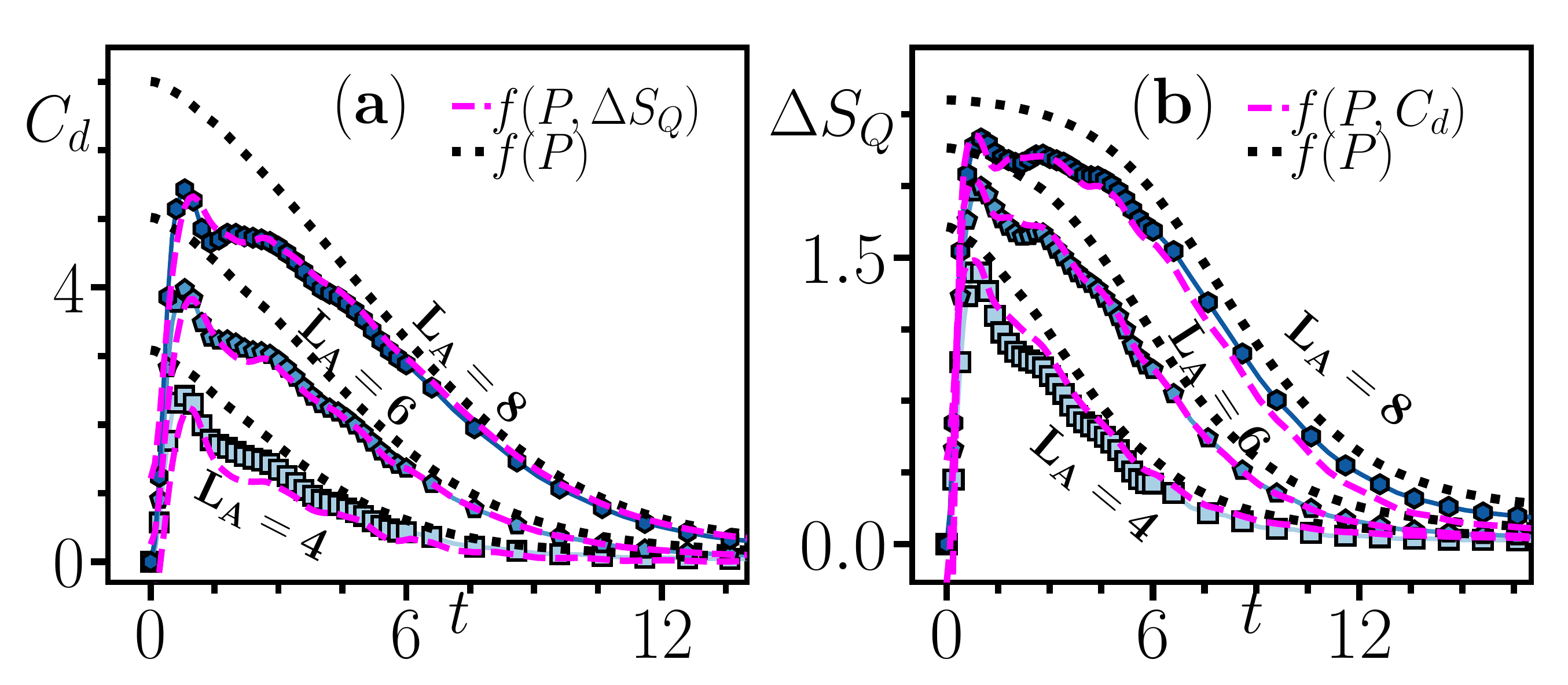}
\caption{\label{fig:mfim}
\textbf{Dynamical equivalence of resources in ergodic Hamiltonian dynamics:}
An identical diagnostics as in Figs.~\ref{fig:rtn} and
\ref{fig:kim} for continuous-time evolution under the MFIM
for \(L=20\) and \(L_A=4,6,8\).
Markers are annealed-averaged data obtained from an exact state vector simulation. Line conventions are identical to Figs. \ref{fig:rtn} and \ref{fig:kim}.}
\end{figure}

Finally, we examine continuous-time, energy-conserving ergodic dynamics of MFIM. Figure~\ref{fig:mfim} displays 
MFIM results with the underlying moments averaged over $200$ random $z$-basis
initial states before taking the logarithm.  
In the case of continuous time evolution,
we evaluate 
$\mathcal M_p\to\{1,\dots,L_A-1\}$ lifting the brickwall parity constraint. Here, the purity-only curves retain the overall decay
scale at late times but do not reproduce the nonmonotonic transient. However, the inclusion of the
first-order construction again remains accurate: $\Delta S_Q$ predicts the
full $C_d$ trajectory, and $C_d$ likewise reconstructs $\Delta S_Q$.
The persistence of this relation despite continuous time, energy
conservation, and additional slow modes demonstrates that the dynamical
equivalence of quantum resources extends beyond the discrete circuit dynamics to all generic ergodic unitary many-body systems.

Additionally, this equivalence is not limited to the pair $(C_d,\Delta S_Q)$ featured
above. As detailed in the End Matter, the position-resolved dipole
asymmetry $\Delta S_P$ can also be reconstructed using a similar strategy across various ergodic dynamics, as shown in
[Fig.~\ref{fig:dipole}]. Moreover, purity-unnormalized quantities, such as the number entropy $S_Q$, can be reconstructed from any purity-normalized witness through a conversion coefficient proportional to the purity $P$ [Fig.~\ref{fig:purity-unnormalized}(a--c)]. Conversely, the $S_Q$-based reconstruction of purity-normalized quantities, for example $C_d$ [Fig.~\ref{fig:purity-unnormalized}(d--f)] again agrees well for the circuit and Floquet dynamics but becomes ill-conditioned under Hamiltonian evolution, as the cross-class factor $P^{-1}$ amplifies small errors or due to neglected higher-wall contributions and hydrodynamic slow modes.

A stronger form of dynamical equivalence emerges between coherence and
imaginarity at the level of quadratic resource measures. As shown in the End Matter [Fig. \ref{fig:imag}], the dephasing and transpose
boundary operators have identical contractions with every configuration
in the local replica sector, and consequently, these two monotones
obey $I(\rho_A(t))
=
-\log_q\!\left[
\frac{1+q^{-C_d(t)}}{2}
\right]$.
Unlike Eq.~\eqref{eq:equivalence_main}, this identity does not rely on
the zero- or one-wall approximation: it holds exactly at each circuit depth,
for arbitrary subsystem size in the Haar-random circuits and remains highly accurate in KIM and MFIM as we show in the End Matter.

\paragraph{Discussion and outlook.}
Our central result is that distinct second-R\'enyi resource measures cease to evolve independently under ergodic many-body dynamics. Although no relation exists between them for a generic quantum state, the ergodic dynamics collapse these measures onto a low-dimensional manifold: the purity fixes the zero-wall contribution, and one additional resource fixes the leading one-wall amplitude. Hence, a target resource can be reconstructed from the purity and a single witness throughout the entire evolution. This reduction is derived for Haar-random circuits and remains to accurately describe also chaotic Floquet and energy-conserving Hamiltonian dynamics, confirming the robustness of our results. 
For Haar-random evolution, we further show that coherence and imaginarity obey the exact relation at all times, which remains remarkably accurate for chaotic Floquet and Hamiltonian settings. Our results provide explicit resource-to-resource relations, identify their microscopic replica-space origin and demonstrate the emergent simplicity of the ergodic evolution that renders the information contained in \textit{different} resource trajectories \textit{dynamically equivalent}.

Our results open several conceptual avenues for exploration. One natural direction is extending the equivalence to higher replicas, e.g., for nonstabilizerness~\cite{leone2022stabilizer,turkeshi2025magic} and fermionic non-Gaussianity~\cite{sierantfermionicmagicresourcesquantum2025,haug2026practicaltestswitnessesfermionic}, which are not captured at the quadratic level.
In these situations, multiple permutation-wall modes may require several witnesses. Determining how conservation laws~\cite{Rakovszky2019hydrodynamics,Khemani2018operatorspreading,Michailidis2024hydrodynamics,aditya2026coherencedynamicsquantummanybody} enlarge the slow replica manifold is another exciting possibility. Applying this framework to integrable~\cite{VidmarRigol2016Gibbsensemble} and ergodicity-breaking~\cite{Abanin2019MBLColloquium,Sierant2025MBLReview,Serbyn2021ergodicitybreaking,Chandran23} dynamics is another compelling direction for future study. Moreover, these relations offer a direct experimental advantage: measuring the purity and one accessible witness through randomized measurements~\cite{Brydges2019Probing,Satzinger2021Topological} or two-copy protocols~\cite{Daley2012Entanglement,Islam2015Entanglement} could enable otherwise costly resource measures to be inferred without full state tomography. We leave these questions for future investigations.

\begin{acknowledgments}
\textit{Acknowledgments.---} S.A acknowledges Sara Murciano, Filiberto Ares, and Pasquale Calabrese for insightful discussions on entanglement asymmetries and related collaborations.
S.A. acknowledges support from Alexander von Humboldt Foundation.
X.T. acknowledges support from DFG Emmy Noether Programme proposal
``Digital Quantum Matter Out-of-Equilibrium'' No. 560726973, DFG under
Germany's Excellence Strategy -- Cluster of Excellence Matter and Light for
Quantum Computing (ML4Q) EXC 2004/2 -- 390534769, and DFG Collaborative
Research Center (CRC) 183 Project No. 277101999 -- project B01.
P.S. acknowledges fellowship within the ``Generaci\'on D'' initiative,
Red.es, Ministerio para la Transformaci\'on Digital y de la Funci\'on
P\'ublica, for talent attraction (C005/24-ED CV1), funded by the European
Union NextGenerationEU funds, through PRTR.
\end{acknowledgments}

\bibliography{ref.bib}

\onecolumngrid 
\section*{End Matter} 
\twocolumngrid 

\section{Exact relation between coherence and imaginarity for $k=2$ annealed averaging }
\label{sec:endmatter_exact_Cd_I}

We now show that there exists an exact relation between the R\'enyi-2 coherence \(C_d\)
and imaginarity \(I(\rho_A)\) under Haar-random circuit evolution. In
contrast to Eq.~\eqref{eq:equivalence_main}, this identity requires
neither a zero-wall nor a one-wall approximation, but rather holds exactly. This thus implies that $C_d$ and $I(\rho_A)$ are not independent quantities under Haar random evolution at the two-replica level. To show that, let us define
$P(t)=\mathbb E_U\,{\rm Tr}\!\left[\rho_A(t)^2\right]$, $
Z_d(t)=\mathbb E_U\,{\rm Tr}\!\left[(\rho_A^D(t))^2\right]$ and $
Z_I(t)=\mathbb E_U\,{\rm Tr}\!\left[\mathfrak R(\rho_A(t))^2\right]$.
The corresponding annealed resource monotones are
\begin{equation}
q^{-C_d(t)}=\frac{Z_d(t)}{P(t)},
\qquad
q^{-I(\rho_A(t))}=\frac{Z_I(t)}{P(t)},
\label{eq:EM_resources}
\end{equation}
where $q$ is the local Hilbert space dimension. On the single-site two-replica space, we define \(e_j\) and \(s_j\) be
the identity and swap configurations. To compute $C_{d}$ and $I(\rho_A)$, one needs to consider the dephasing and transpose
boundaries locally, i.e.,
\begin{equation}
W_j=\sum_{a=0}^{q-1}|aa\rangle\langle aa|,
\qquad
T_j=s_j^\Gamma
=\sum_{a,b=0}^{q-1}|aa\rangle\langle bb|,
\label{eq:EM_local_boundaries}
\end{equation}
where \(\Gamma\) is the partial transpose on one replica. Interestingly, both the boundary
contractions at \(k=2\) permutation basis are locally
identical, i.e.,
\begin{equation}
{\rm Tr}(W_j e_j)
={\rm Tr}(W_j s_j)
={\rm Tr}(T_j e_j)
={\rm Tr}(T_j s_j)
=q.
\label{eq:EM_equal_rows}
\end{equation}

Therefore, for a subsystem $A$ of $L_{A}$ sites, writing
$W_A=\bigotimes_{j\in A}W_j$ and 
$T_A=\bigotimes_{j\in A}T_j$,
Eq.~\eqref{eq:EM_equal_rows} factorizes site by site. Hence, for every
replica configuration
\(\boldsymbol{\sigma}=\bigotimes_{j\in A}\sigma_j\), with
\(\sigma_j\in\{e_j,s_j\}\),
\begin{equation}
{\rm Tr}(W_A\boldsymbol{\sigma})
=
{\rm Tr}(T_A\boldsymbol{\sigma}).
\label{eq:EM_configuration}
\end{equation}

Now, the $k=2$ Haar replicated state belongs exactly to this permutation
sector, which can be written as
\begin{equation}
X_t=\mathbb E_U[\rho_A(t)^{\otimes2}]
=
\sum_{\boldsymbol{\sigma}\in\{e,s\}^{\otimes L_A}}
w_t(\boldsymbol{\sigma})\,\boldsymbol{\sigma},
\label{eq:EM_replica_expansion}
\end{equation}
where the sum contains configurations with arbitrary numbers of domain walls. Eqs.~\eqref{eq:EM_configuration} and
\eqref{eq:EM_replica_expansion} thus imply, without any approximation,
\begin{align}
\mathbb E_U\,{\rm Tr}(\rho_A\rho_A^T)
=\mathbb E_U\,{\rm Tr}[(\rho_A^D)^2]
=Z_d(t).
\label{eq:EM_transpose_dephasing}
\end{align}

\begin{figure*}[htb]
\centering
\includegraphics[width=0.85\textwidth]{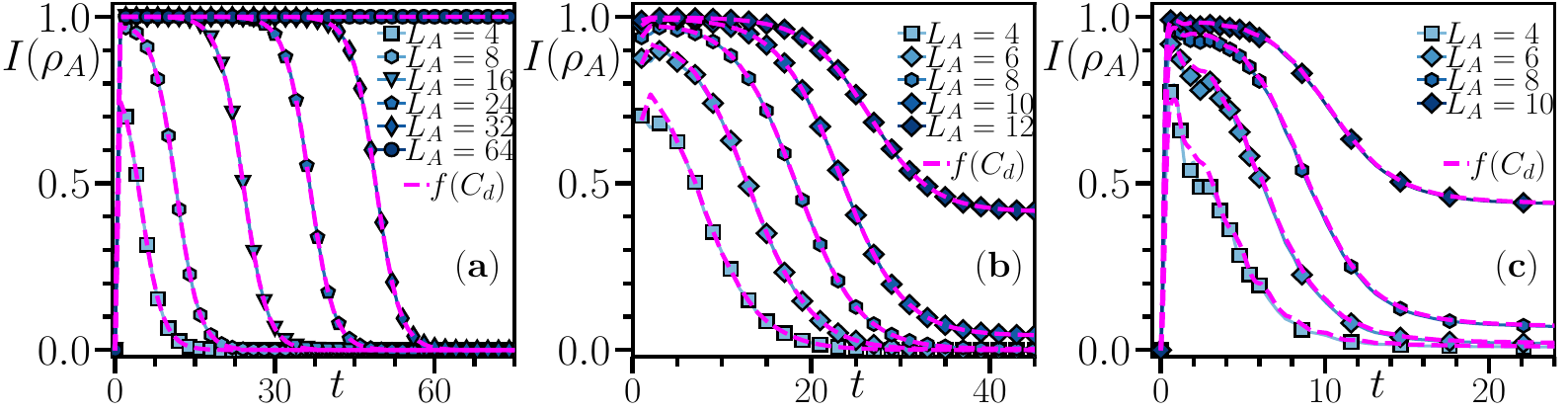}
\caption{\textbf{$C_d$ and $I(\rho_A)$ relation across ergodic dynamics:} Test of the relation between $C_d$ and $I(\rho_A)$ in (a) Haar-random circuits ($L=256$), (b) the Floquet kicked Ising model ($L=24$), and (c) ergodic Hamiltonian dynamics ($L=20$), for several subsystem sizes ($L_A$). In all three panels, markers denote the numerically obtained $I(\rho_A)$, while dotted lines show its prediction from $C_d$. The relation is exact for the Haar-random circuit in (a) and remains remarkably accurate for the Floquet and Hamiltonian dynamics in (b) and (c), respectively.}
\label{fig:imag}
\end{figure*}

Finally, Hermiticity of \(\rho_A\) gives
\begin{equation}
{\rm Tr}[\mathfrak R(\rho_A)^2]
=
\frac12\left[
{\rm Tr}(\rho_A^2)
+
{\rm Tr}(\rho_A\rho_A^T)
\right]
\label{eq:EM_realification}
\end{equation}
Taking the ensemble average and using
Eq.~\eqref{eq:EM_transpose_dephasing}, we obtain
\begin{equation}
Z_I(t)=\frac12\left[P(t)+Z_d(t)\right],
\label{eq:EM_moment_lock}
\end{equation}
which dividing by \(P(t)\) and applying Eq.~\eqref{eq:EM_resources} yields
\begin{eqnarray}
q^{-I(\rho_A(t))}
&=&\frac12\left[1+q^{-C_d(t)}\right],\nonumber\\
I(\rho_A(t))
&=&
-\log_q\!\left[
\frac{1+q^{-C_d(t)}}{2}
\right]. 
\label{eq:EM_exact_lock}
\end{eqnarray}
Equation~\eqref{eq:EM_exact_lock} holds at every circuit depth for
which the Haar-averaged state lies in the local \(k=2\) permutation
sector, for every subsystem size. This is therefore
an exact identity between annealed second moments, rather than a
first-order resource-equivalence~\eqref{eq:equivalence_main}. Further, for \(q=2\), Eq.~\eqref{eq:EM_exact_lock} becomes $I(\rho_A)
=
1-\log_2\!\left(1+2^{-C_d}\right)$, and
since \(C_d\geq0\), it follows that $0\leq I(\rho_A)<1$, and
$I(\rho_A)\rightarrow1$ as
$C_d\rightarrow\infty$.
Thus, \(I=1\) is the maximum saturation value for qubits. More generally, for local
dimension \(q\), $I(\rho_A)\leq\log_q2$. We now turn to examine this relation across different classes of ergodic dynamics, as shown in Fig.~\ref{fig:imag}. For Haar-random dynamics, the relation is exact, as demonstrated in Fig.~\ref{fig:imag}(a). Remarkably, it remains highly accurate beyond the Haar setting, both for the Floquet kicked Ising chain [Fig.~\ref{fig:imag}(b)] and for continuous-time ergodic Hamiltonian dynamics with GUE spectral statistics [Fig.~\ref{fig:imag}(c)]. Taken together, these results suggest that $C_d$ and $I(\rho_A)$ are not independent under ergodic dynamics, but instead become dynamically locked by an emergent relation that persists across distinct ergodic settings.

\section{The prediction of Dipole asymmetry from other resources under ergodic dynamics}
\label{sec:endmatter_dipole}

Here we extend the analysis of the main text to the dipole asymmetry
$\Delta S_P$. In contrast to the charge asymmetry, its
boundary operator depends on the positions of the swap sites~\cite{SM}. Hence, $\Delta S_P$ resolves the
spatial structure of the replicated state. Nevertheless, its annealed
second moment remains a linear function of the same replicated state
$X_t$, and the reduction of
Eqs.~\eqref{eq:purity_law_main} and \eqref{eq:equivalence_main} can be again applied, with $\mathcal N_{\Delta S_P}$ evaluated from
the position-resolved manner~\cite{SM}.

\begin{figure*}[htb]
\centering
\includegraphics[width=0.84\textwidth]{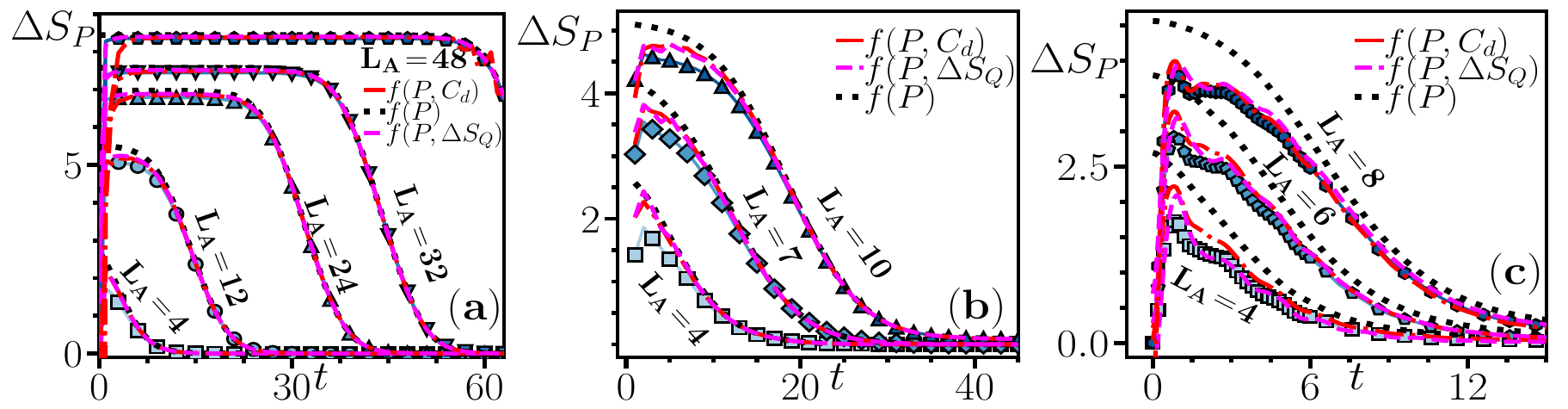}
\caption{\textbf{Dynamical equivalence of $\Delta S_{p}$ under ergodic dynamics.} Dipole asymmetry $\Delta S_{p}$ and its predictions from the purity $P$, coherence $C_d$, and charge asymmetry $\Delta S_{Q}$ under (a) Haar-random circuit dynamics ($L=256$), (b) Floquet kicked Ising evolution ($L=24$), and (c) continuous-time mixed-field Ising evolution ($L=20$). Markers denote the numerical data for various $L_A$'s. The black dash-dotted line showcases the purity-only prediction, while the red and magenta lines illustrate the predictions obtained from $(P,C_d)$ and $(P,\Delta S_{Q})$, respectively.}
\label{fig:dipole}
\end{figure*}

In Figure~\ref{fig:dipole}, we showcase the predictions across the three
ergodic settings: (a)~the Haar-random brick-wall circuit ($L=256$),
(b)~the Floquet kicked Ising model ($L=24$), and (c)~the continuous-time
mixed-field Ising model ($L=20$) for various $L_A$'s. In all cases the purity-only
reconstruction (black dash-dotted) captures the late-time relaxation
envelope, however, it fails to capture the early-time behavior (except when there exist early-time plateaus), most visibly for the Floquet and Hamiltonian dynamics. The inclusion of
one-wall amplitude through either witness---the coherence $C_d$ (red) or
the charge asymmetry $\Delta S_Q$ (magenta)---nevertheless shows significant improvement. Both reconstructions closely follow the numerical data, including the initial
rise, the $L_A$-dependent maxima, and the subsequent relaxation. The dynamical
equivalence thus encompasses position-resolved asymmetries across all
ergodic dynamics.

\section{Purity-unnormalized resources under ergodic dynamics}
\label{sec:endmatter_unnormalized}

\begin{figure*}[htb]
\includegraphics[width=0.8\textwidth]{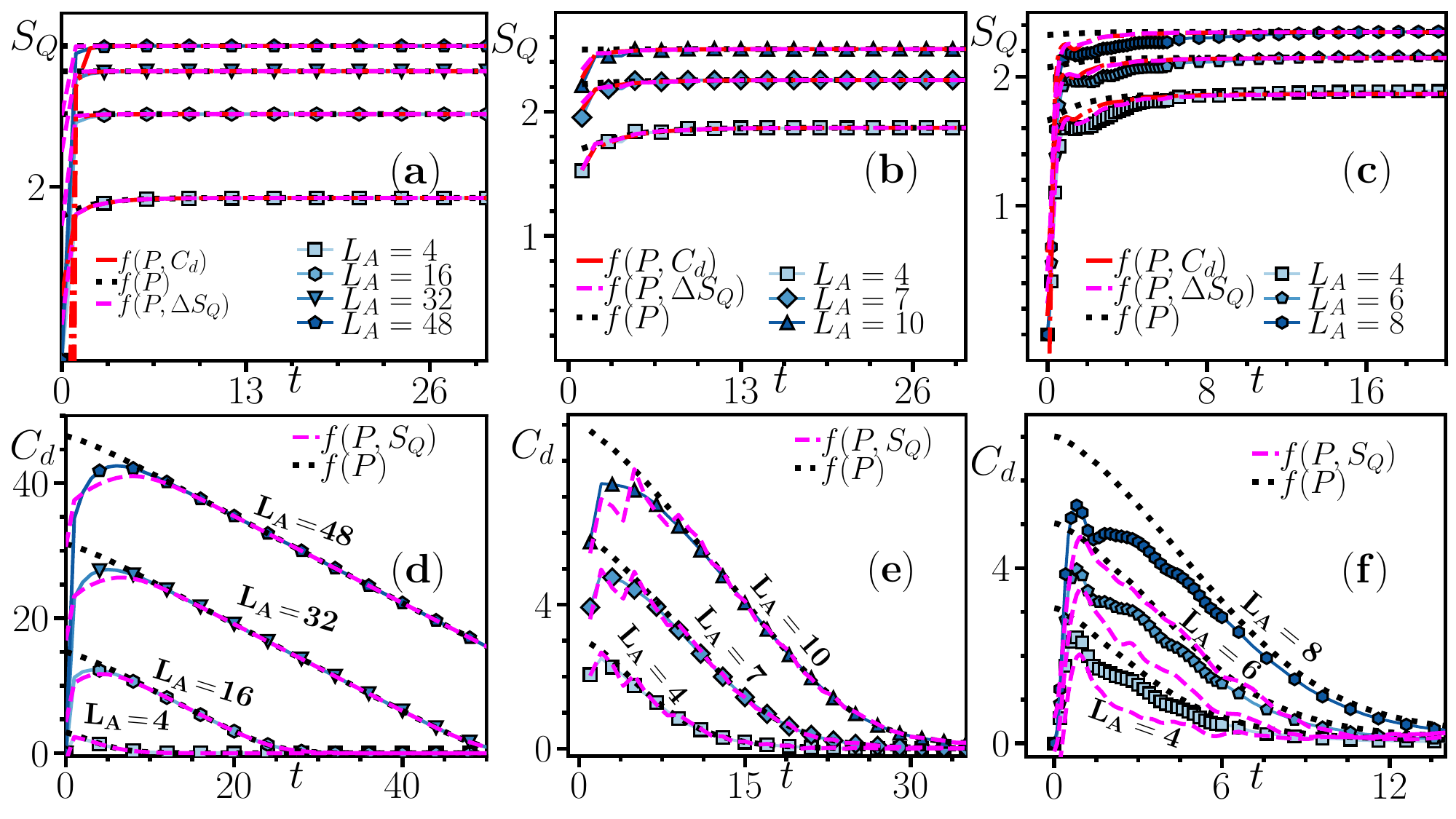}
\caption{\textbf{Dynamical equivalence of number entropy $S_Q$ across ergodic dynamics.}
(a--c) Number entropy $S_Q$ reconstructed from the purity $P$ and different purity-normalized resources under Haar-random circuit dynamics ($L=256$), Floquet kicked Ising evolution ($L=24$), and mixed-field Ising Hamiltonian dynamics ($L=20$), respectively. (d--f) Conversely, the purity-normalized resource $C_d$ reconstructed from $S_Q$ for the same three ergodic dynamics as in panels (a--c). In all panels, markers denote the numerical data, while the black dash-dotted lines show the purity-only predictions. In panels (a--c), the red and magenta lines denote the reconstructions from $(P,C_d)$ and $(P,\Delta S_Q)$, respectively. In panels (d--f), the magenta dash-dotted lines show the $S_Q$-based reconstruction of $C_d$.}
\label{fig:purity-unnormalized}
\end{figure*}

Finally, we turn to purity-unnormalized quantities, focusing on the number entropy $S_Q$; the charge variance $\operatorname{Var}(Q_A)$ and their dipole counterparts exhibit qualitatively identical behavior. Since $\eta_{S_Q}=1$, the conversion coefficient relating $S_Q$ to a purity-normalized witness contains an explicit factor of the purity, $\rho_{S_QR'}\propto P$, in Eq.~\eqref{eq:equivalence_main}. Figure~\ref{fig:purity-unnormalized}(a--c) shows $S_Q$ reconstructed from $(P,C_d)$ (red) and $(P,\Delta S_Q)$ (magenta) for the Haar-random circuit, the Floquet kicked Ising model, and the mixed-field Ising Hamiltonian, respectively. In all three settings, the first-order reconstruction captures better actual numerical data, including the early-time regime in which the purity-only prediction showcase slight deviation, especially in case of ergodic Hamiltonian dynamics.

In addition, Figure~\ref{fig:purity-unnormalized}(d--f) examines the reverse reconstruction, with the purity-unnormalized quantity $S_Q$ serving as the witness for the purity-normalized target $C_d$. The reconstruction agrees well with the numerical data for the circuit (d) and Floquet (e) dynamics, whereas visible rapid deviations emerge under Hamiltonian evolution (f). This can cause due to the cross-class coefficient $\rho_{C_dS_Q}\propto P^{-1}$ amplifies small errors in the prediction, whether numerical or due to neglected higher-wall and hydrodynamic slow modes by a factor of order $q^{L_A}$ as thermalization drives $P\to D^{-1}$. In all cases, sole-purity prediction fails to capture the early-time growth regime before relaxation.

\widetext
\clearpage
\begin{center}
{\Large \textbf{Supplemental Material for\\[2pt]
\titleinfo}}\\[6pt]
Sreemayee Aditya,
Xhek Turkeshi,
Piotr Sierant
\end{center}

\setcounter{equation}{0}
\setcounter{figure}{0}
\setcounter{table}{0}
\setcounter{page}{1}
\renewcommand{\theequation}{S\arabic{equation}}
\setcounter{figure}{0}
\renewcommand{\thefigure}{S\arabic{figure}}
\renewcommand{\thepage}{S\arabic{page}}
\renewcommand{\thesection}{S\arabic{section}}
\renewcommand{\thetable}{S\arabic{table}}
\makeatletter

\renewcommand{\thesection}{\arabic{section}}
\renewcommand{\thesubsection}{\thesection.\arabic{subsection}}
\renewcommand{\thesubsubsection}{\thesubsection.\arabic{subsubsection}}

\vspace{1em}

\setcounter{section}{0}
\renewcommand{\thesection}{S\arabic{section}}
\renewcommand{\theequation}{S\arabic{equation}}
\renewcommand{\thefigure}{S\arabic{figure}}
\renewcommand{\thetable}{S\arabic{table}}
\setcounter{equation}{0}
\setcounter{figure}{0}
\setcounter{table}{0}

In this Supplemental Material, we provide a detailed derivation of the
resource-equivalence relations stated in the main text for Haar random circuits. The argument
proceeds in four steps: (i) Haar averaging at the two-replica level
restricts the dynamics to local identity and swap configurations;
(ii) normalization and the measured purity fix the zero-wall component;
(iii) the leading correction is carried by a single spectral one-wall
mode; (iv) eliminating the common amplitude of this mode relates any two
resource diagnostics. The four sections below follow these steps in
order.

As in the main text, we consider a chain of $L$ qubits, with local dimension $q=2$, initialized
in $\ket{0\cdots0}$ and evolved by a Haar-random brick-wall circuit. The
boundary subsystem $A$ contains the first $L_A$ sites and has dimension
$D=q^{L_A}$. Further, we denote by $e_j$ and $s_j$ the identity and swap configurations in
the doubled Hilbert space of site $j$, so that $\Tr e_j=q^2$ and $\Tr s_j=q$. All diagnostics of interest are annealed second-R\'enyi quantities, and
each is a linear function (also referred to as readout) of the same replicated state,
\begin{equation}
\mathcal Z_R(t)=\Tr(B_RX_t)
=\mathbb E_U\Tr\!\left[B_R\rho_A(t)^{\otimes2}\right].
\label{eq:SM_boundary_moment}
\end{equation}
The measured observables are normalized by the purity for purity-normalized resources, whereas for the remaining resources it enters in an unnormalized manner. To treat both cases on equal footing, we introduce the uniform
notation
\begin{equation}
Y_R(t)=\frac{\mathcal Z_R(t)}{\eta_R(t)},
\qquad
\eta_R(t)=
\begin{cases}
P(t),&R\in{\text{ Purity-normalized}},\\
1,&R\in{\text{Purity-unnormalized}} .
\end{cases}
\label{eq:SM_normalized_readout}
\end{equation}
For a two-site Haar random gate $u\in U(q^2)$, Schur--Weyl duality implies that
\begin{equation}
\Phi(O)=\mathbb E_u\!\left[
u^{\otimes2}O(u^\dagger)^{\otimes2}\right]
\label{eq:SM_Haar_channel}
\end{equation}
projects onto the span of the two-site identity and swap. Matching the
two invariant overlaps gives the local rule
$ee\mapsto ee$, $ss\mapsto ss$, and 
$es,\;se\mapsto r(ee+ss)$, where $r=\frac{q}{q^2+1}$. Consequently, after one complete layer, the replicated state reduced to $A$ can be recast as
\begin{equation}
X_t=\sum_{\sigma_A\in\{e,s\}^{L_A}}
v_t[\sigma_A]\bigotimes_{x\in A}\sigma_x.
\label{eq:SM_configuration}
\end{equation}
All diagnostics therefore probe the same coefficients $v_t$ and differ
only in their boundary operator contractions; the remainder of the proof exploits
this common structure.
\section{Zero-wall manifold}
We begin with the sector that dominates at late times. The uniform
configurations span the zero-wall sector,
\begin{equation}
X_t^{(0)}=a_e e_A+a_s s_A.
\label{eq:SM_zero_wall}
\end{equation}
Normalization and the measured purity give
\begin{equation}
D^2a_e+Da_s=1,\qquad Da_e+D^2a_s=P,
\end{equation}
which leads to $a_e=\frac{D-P}{D(D^2-1)}$, and
$a_s=\frac{DP-1}{D(D^2-1)}$, in agreement with the main text.
Defining the two uniform boundary contractions
$n_e^R=\Tr(B_Re_A)$ and $n_s^R=\Tr(B_Rs_A)$,
one obtains the purity-controlled law
\begin{equation}
\mathcal Z_R^{(0)}(P)
=\frac{(D-P)n_e^R+(DP-1)n_s^R}{D(D^2-1)}.
\label{eq:SM_general_purity_law}
\end{equation}
For the purity-normalized resources, $n_e^R=D$ and
$Y_R=q^{-M_R}$, so that for qubits
\begin{equation}
2^{-M_R^{(0)}}=\frac{D^2-n_s^R}{PD(D^2-1)}+\frac{n_s^{R}-1}{D^2-1}.
\label{eq:SM_classA_purity_law}
\end{equation}
For the purity-unnormalized resources, i.e., for number entropy and charge variance, the same contraction
gives
\begin{equation}
\begin{aligned}
2^{-S_Q^{(0)}}
&=\frac{\sigma(D-P)+D(DP-1)}{D(D^2-1)},\\
\left\langle{\rm Var}\,Q_A\right\rangle^{(0)}
&=\frac{L_A D(D-P)}{4(D^2-1)},
\end{aligned}
\qquad
\sigma=\binom{2L_A}{L_A}.
\label{eq:SM_classB_purity_laws}
\end{equation}
At this order every diagnostic is thus pinned by the single measured
number $P(t)$; the rest of the argument quantifies the leading correction
to this statement.

\section{Leading spectral correction}
Let us now turn to the leading correction beyond the zero-wall sector, which is carried by the one-wall mode. A Haar gate leaves aligned pairs, $ee$ or $ss$, unchanged, while a mixed pair, $es$ or $se$, is mapped to $ee$ or $ss$. Consequently, a wall can move, annihilate with a neighboring wall, or be absorbed at an open boundary, but cannot generate additional walls. Hence
$\Pi_m\mathcal T\Pi_n=0$ for $m>n$,
where $\Pi_n$ projects onto the $n$-wall sector, so that $\mathcal T$ is triangular in wall number: its diagonal blocks conserve wall number, whereas off-diagonal blocks connect only to sectors with fewer walls.

After each brickwork layer, walls reside on bonds of fixed parity $b\equiv p=t\bmod 2$. In the diagonal one-wall block, a wall hops to either neighboring bond with weight $r$, giving
\begin{equation}
\mu_{t+1}(b)=r[\mu_t(b-1)+\mu_t(b+1)],
\qquad
\mu_t(-1)=\mu_t(L-1)=0 .
\label{eq:SM_wall_walk}
\end{equation}
The absorbing boundaries account for a wall leaving the chain and entering the zero-wall sector. The eigenmodes are therefore
$\phi_k(b)=\sin\frac{\pi k(b+1)}{L}$, and $\lambda_k=2r\cos\frac{\pi k}{L}$, where $k=1,\ldots,L-1$,
so the dominant one-wall mode has $\lambda_1=2r\cos(\pi/L)$ and profile $\phi_1$.
Off-diagonal blocks dress this mode with lower-wall components. We denote the resulting eigenvector at parity $p$ by $\widetilde V_{1,p}$. After tracing out $B$, we remove its component along the zero-wall manifold and define
\begin{equation}
\mathcal{Q}_p=\Pi_\perp\Tr_B\widetilde V_{1,p},
\qquad
\Tr \mathcal{Q}_p=\Tr(s_A\mathcal{Q}_p)=0.
\label{eq:SM_spectral_mode}
\end{equation}
Thus the leading correction changes neither the normalization nor the
measured purity. Provided the initial state has nonzero overlap with this
mode, the reduced replicated state takes the asymptotic form
\begin{equation}
X_t=X_t^{(0)}[P(t)]+\mathcal A_1(t)\mathcal{Q}_p.
\label{eq:SM_spectral_expansion}
\end{equation}

The nonuniform part of $\Tr_B\widetilde V_{1,p}$ arises only when the
wall lies inside $A$, namely on
$0\le b\le L_A-2$. A wall lying in $B$ leaves $A$ uniformly $e$ or
$s$ and is removed by $\Pi_\perp$. For an interior wall there are two
orientations,
\[
s^{b+1}e^{L_A-1-b},
\qquad
e^{b+1}s^{L_A-1-b}.
\]
They have equal amplitudes before tracing $B$, by the global
$e\leftrightarrow s$ symmetry of the gate action. Their bath
traces, however, differ. In the first orientation the
outermost block of $A$ is $e$, so continuity extends the $e$ domain
through $B$ and gives $q^{2(L-L_A)}$. In the second, it is $s$, giving
$q^{L-L_A}$. After absorbing the common first factor into
$\mathcal A_1(t)$, their relative weight thus becomes
$\kappa=q^{-(L-L_A)}$.
Approximating the one-wall component of the dressed mode by the bare dominant
sine profile $\phi_1$ therefore gives, up to an irrelevant overall
normalization,
\begin{equation}
\begin{aligned}
\mathcal{Q}_p^{\rm bare}
=\Pi_\perp\!
\sum_{\substack{0\le b\le L_A-2\\b\equiv p\ ({\rm mod}\ 2)}}
\sin\!\left[\frac{\pi(b+1)}{L}\right]
\Big[
&\ket{s^{b+1}e^{L_A-1-b}}
+\kappa\ket{e^{b+1}s^{L_A-1-b}}
\Big],
\end{aligned}
\label{eq:SM_Mp}
\end{equation}
where, as above, $\kappa=q^{-(L-L_A)}$ is the relative bath-trace weight of the
two wall orientations. To make the projection explicit, we denote the
unprojected sum in Eq.~\eqref{eq:SM_Mp} by $\widehat {\mathcal Q_p}$ and write
\[
\mathcal{Q}_p^{\rm bare}=\widehat {\mathcal{Q}_p}+\delta a\,e_A+\delta b\,s_A .
\]
The two added terms contain no wall; they merely subtract the components
already fixed by normalization and purity.

To fix the projection,
we denote $S_e\equiv\Tr\widehat {\mathcal{Q}_p}$, and
$S_s\equiv\Tr(s_A\widehat {\mathcal{Q}_p})$.
Since $\Tr e_A=D^2$, $\Tr s_A=D$, and
$\Tr(s_A^2)=D^2$, the orthogonality conditions in
Eq.~\eqref{eq:SM_spectral_mode} become
\[
\begin{pmatrix}
D^2&D\\ D&D^2
\end{pmatrix}
\begin{pmatrix}\delta a\\ \delta b\end{pmatrix}
=-
\begin{pmatrix}S_e\\ S_s\end{pmatrix},
\]
leading to
$\delta a=\frac{S_s-DS_e}{D(D^2-1)}$ and
$\delta b=\frac{S_e-DS_s}{D(D^2-1)}$. The exact leading spectral mode can contain additional dressing inherited from the lower-wall blocks of the
triangular transfer matrix. In what follows, we retain its bare one-wall approximation,
$\mathcal{Q}_{p}\simeq \mathcal{Q}_{p}^{\text{bare}}$, $\mathcal N_R \simeq \mathcal N_{R}^{\text{bare}}$, and henceforth suppress the superscript ``bare". The projected mode therefore changes
neither $\Tr X_t$ nor the measured purity, and every remaining boundary
signal is strictly independent of the zero-wall contribution. With the
mode in hand, we can now extract the resource-equivalence relation.

\section{Elimination of the common amplitude}
Contracting Eq.~\eqref{eq:SM_spectral_expansion} with $B_R$ gives
\begin{equation}
\mathcal Z_R(t)-\mathcal Z_R^{(0)}[P(t)]
\simeq\mathcal A_1(t)\mathcal N_R,
\qquad
\mathcal N_R=\Tr(B_R\mathcal{Q}_p).
\label{eq:SM_factorization}
\end{equation}
The crucial observation is that the same amplitude $\mathcal A_1(t)$ multiplies the boundary sensitivity $\mathcal N_R$ for every diagnostic. Writing $\delta\mathcal Z_R
\equiv\mathcal Z_R-\mathcal Z_R^{(0)}$ and $\delta\mathcal Z_{R'}
\equiv\mathcal Z_{R'}-\mathcal Z_{R'}^{(0)}$,
Eq.~\eqref{eq:SM_factorization} gives
$\delta\mathcal Z_R=\mathcal A_1\mathcal N_R+\epsilon_R$ and
$\delta\mathcal Z_{R'}=\mathcal A_1\mathcal N_{R'}+\epsilon_{R'}$, where
$\epsilon_R$ collects the subleading spectral corrections. Solving the
witness relation for $\mathcal A_1$ and substituting into the target
relation yields
\begin{equation}
\begin{aligned}
\mathcal Z_R-\mathcal Z_R^{(0)}
\simeq&\frac{\mathcal N_R}{\mathcal N_{R'}}
\left(\mathcal Z_{R'}-\mathcal Z_{R'}^{(0)}\right).
\label{eq:resourceconversion}
\end{aligned}
\end{equation}
Finally, using $\mathcal Z_R=\eta_RY_R$ and
$\mathcal Z_R^{(0)}=\eta_RY_R^{(0)}$ converts
Eq.~\eqref{eq:resourceconversion} into the improved law of the main text,
\begin{equation}
Y_R(t)\simeq Y_R^{(0)}[P(t)]
+\rho_{RR'}[P(t)]\left\{Y_{R'}(t)-Y_{R'}^{(0)}[P(t)]\right\},
\qquad
\rho_{RR'}=\frac{\eta_{R'}}{\eta_R}\,
\frac{\mathcal N_R}{\mathcal N_{R'}}.
\label{eq:SM_improvedlaw_final}
\end{equation}
For two resources of the same class the normalization factors cancel and
$\rho_{RR'}$ is a pure number; across classes it carries one factor of
$P^{\pm1}$. All that remains is to evaluate the sensitivities
$\mathcal N_R$, which we do in the final section.

\section{Boundary dictionaries and conversion coefficients}
\label{sec:coefficients}

Our final task is to collect the boundary data needed to evaluate
$\mathcal N_R$. The dictionary of $R$ is
\begin{equation}
c_R[\sigma_A]=
\Tr\!\left(B_R\bigotimes_{x\in A}\sigma_x\right),
\qquad
\sigma_A\in\{e,s\}^{L_A},
\label{eq:SM_dictdef}
\end{equation}
so that $\mathcal Z_R=\sum_{\sigma_A}v_t[\sigma_A]c_R[\sigma_A]$.
Except for dipole-resolved quantities, the dictionary depends only on
the number $m$ of swap sites. The required results follow directly from
the local phase or flux contractions and are summarized in
Table~\ref{tab:SM_dict}. Three shorthands appearing there deserve
definition. For a set of positions $X\subseteq\{1,\dots,L_A\}$,
$N_X(d)$ denotes the number of subsets of $X$ whose position sum equals
$d$; in the dipole rows, $S$ ($E$) is the set of $s$-site ($e$-site)
positions of the configuration at hand, and
$\sigma_{\rm dip}=\sum_dN_{\{1..L_A\}}(d)^2$ is the corresponding
uniform value. The function $T_2(m)$ is the second charge moment of an
$m$-swap configuration,
$T_2(m)=\Tr\big[(Q_A\otimes Q_A)\,s^me^{L_A-m}\big]$; expanding
$Q_A\otimes Q_A=\sum_{x,y}\hat n_x\otimes\hat n_y$ into per-site
contractions gives, for $q=2$ and with $k=L_A-m$,
\begin{equation}
T_2(m)=k\,4^{k-1}2^m+m\,4^k2^{m-1}+k(k-1)4^{k-1}2^m
+4km\,4^{k-1}2^{m-1}+m(m-1)4^k2^{m-2}.
\label{eq:SM_T2}
\end{equation}

\begin{table}[t]
\caption{\label{tab:SM_dict}
Dictionary summary. The first six rows depend on the configuration only
through the number $m$ of swap sites; the dipole rows also depend on the
positions of the $s$-/$e$-sites. Here $\sigma=\binom{2L_A}{L_A}$ and
$\sigma_{\rm dip}=\sum_d N_{\{1..L_A\}}(d)^2$.}
\begin{ruledtabular}
\begin{tabular}{lccc}
quantity & $c_R[\sigma_A]$ & $n_e$ & $n_s$\\
\hline
$C_d$ & $D$ & $D$ & $D$\\
$I$ & $\frac12(q^{L_A+m}+D)$ & $D$ & $\frac{D(D+1)}{2}$\\
$\Delta S_{\mathbb Z_2}$ & $\frac12(q^{L_A+m}+D\delta_{m,0})$ & $D$ & $\frac{D^2}{2}$\\
$\Delta S_Q$ & $q^{L_A-m}\binom{2m}{m}$ & $D$ & $\sigma$\\
$S_Q$ & $q^m\binom{2(L_A-m)}{L_A-m}$ & $\sigma$ & $D$\\
$\Delta S_P$ & $q^{L_A-m}\sum_dN_S(d)^2$ & $D$ & $\sigma_{\rm dip}$\\
$S_P$ & $q^m\sum_dN_E(d)^2$ & $\sigma_{\rm dip}$ & $D$\\
${\rm Var}\,Q_A$ & $-T_2(m)$ & $-\frac{L_A^2D^2}{4}$
& $-\frac{DL_A(L_A+1)}{4}$
\end{tabular}
\end{ruledtabular}
\end{table}
For the remaining coefficients, all dependence on the one-wall position
is contained in
\begin{equation}
\mathfrak G_p(x):=\sum_{m\in\mathcal \mathcal{M}_p}\sin\frac{\pi m}{L}\,x^m
={\rm Im}\!\left[z^{m_0}\,\frac{1-z^{2M}}{1-z^2}\right],
\qquad z=x\,e^{i\pi/L},
\label{eq:SM_profilesum}
\end{equation}
where their positions
\begin{equation}
\mathcal \mathcal{M}_p=
\{1\le m\le L_A-1: m\equiv p+1\ ({\rm mod}\ 2)\},
\quad
z=xe^{i\pi/L},
\quad
m_0=\min\mathcal \mathcal{M}_p,
\quad
M=|\mathcal \mathcal{M}_p|.
\end{equation}
We now evaluate the sensitivity
$\mathcal N_R=\Tr(B_R\mathcal{Q}_p)$ explicitly. A common overall
normalization of $\mathcal{Q}_p$ is immaterial because it cancels from
$\mathcal N_R/\mathcal N_{R'}$; we choose the normalization $1/D$ used
below. 
Setting $m=b+1$, the first orientation in Eq.~\eqref{eq:SM_Mp} is
$s^me^{L_A-m}$, containing $m$ swaps, and its direct dictionary
contraction contributes
$\frac1D\sum_{m\in\mathcal \mathcal{M}_p}
\sin\frac{\pi m}{L}\,c_R(m)$.
Its counterweight response follows from the raw traces, which factorize
site by site: using
$\Tr e=q^2$, $\Tr s=q$, $\Tr(se)=q$, and
$\Tr(s^2)=q^2$, one finds
\[
\begin{aligned}
\Tr(s^me^{L_A-m})
&=q^m(q^2)^{L_A-m}=D^2q^{-m},\\
\Tr[s_A(s^me^{L_A-m})]
&=(q^2)^mq^{L_A-m}=Dq^m.
\end{aligned}
\]
After summing over the sine profile and including the
overall $1/D$ normalization, the two traces entering the counterweights
thus become
\[
S_e^{(+)}=D\,\mathfrak G_p(q^{-1}),
\qquad
S_s^{(+)}=\mathfrak G_p(q).
\]

Adding the direct and counterweight terms gives the
single-orientation form factor
\begin{equation}
\mathcal F_R=\frac1D\sum_{m\in\mathcal \mathcal{M}_p}\sin\frac{\pi m}{L}\,c_R(m)
+\frac{\mathfrak G_p(q)\,(n_e-Dn_s)}{D(D^2-1)}
+\frac{\mathfrak G_p(1/q)\,(n_s-Dn_e)}{D^2-1},
\label{eq:SM_formfactor}
\end{equation}
where the superscript $R$ on $(n_e,n_s)$ is suppressed. The second
orientation in Eq.~\eqref{eq:SM_Mp} at the same bond is
$e^ms^{L_A-m}$, containing $L_A-m$ swaps; its contribution follows
from the identical computation applied to the transposed diagnostic,
defined by
\begin{equation}
c_{\widetilde R}(m)=c_R(L_A-m),
\qquad
(n_e^{\widetilde R},n_s^{\widetilde R})=(n_s^R,n_e^R).
\end{equation}

\begin{figure}[!htb]
\centering
\includegraphics[width=0.98\textwidth]{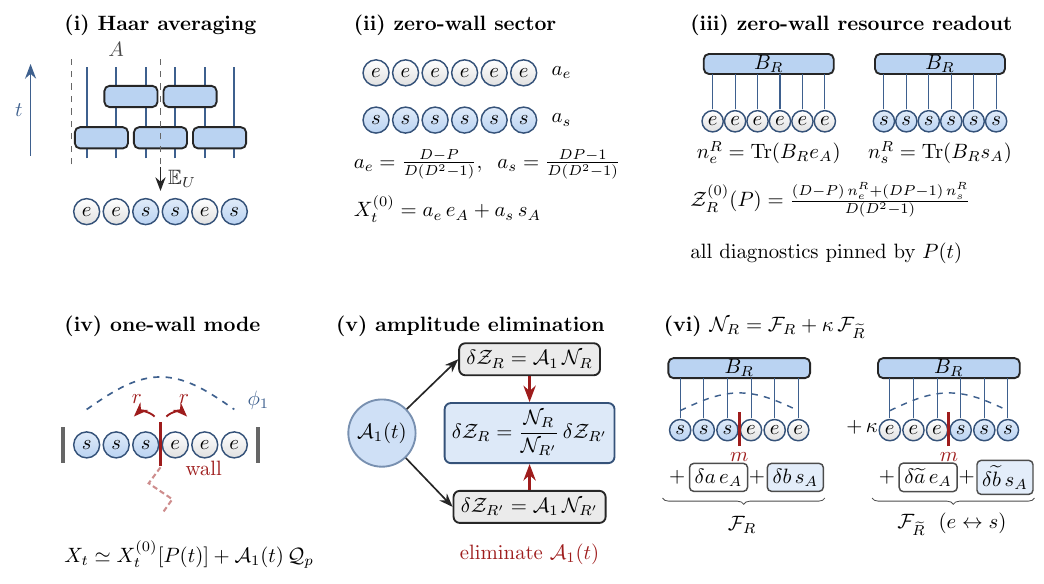}
\caption{\label{fig:SM_steps}Schematic overview of the
proof. (i)~Haar averaging at the two-replica level reduces the brick-wall
dynamics to configurations of local identities $e$ and swaps $s$.
(ii)~The uniform (zero-wall) configurations dominate at late times;
their coefficients $a_e$, $a_s$ are fixed by normalization and the
measured purity $P(t)$. (iii)~Contracting the boundary readout $B_R$
with the uniform configurations $e_A$ and $s_A$ yields the boundary
contractions $n_e^R$, $n_s^R$ and the purity-controlled zero-wall law
$\mathcal Z_R^{(0)}(P)$, which pins every diagnostic to the single
measured number $P(t)$. (iv)~The leading correction is a single one-wall
mode: a domain wall performs a random walk with hopping weight $r$
between absorbing boundaries (thick gray bars), with dominant profile
$\phi_1$ and eigenvalue $\lambda_1$. (v)~The common amplitude
$\mathcal A_1(t)$ multiplies every boundary sensitivity $\mathcal N_R$;
eliminating it between a target $R$ and a witness $R'$ yields the
resource-equivalence relation. (vi)~The boundary sensitivity
$\mathcal N_R=\Tr(B_R\,\mathcal{Q}_p)$: for each wall orientation, the
one-wall sine profile is contracted with the boundary readout $B_R$,
giving the direct term $\frac1D\sum_m\sin\frac{\pi m}{L}\,c_R(m)$, and
is supplemented by its own zero-wall counterweights
($\delta a\,e_A+\delta b\,s_A$ for $s^me^{L_A-m}$, and
$\delta\widetilde a\,e_A+\delta\widetilde b\,s_A$ for the transposed
configuration $e^ms^{L_A-m}$), which generate the correction terms
containing $\mathfrak G_p(q)$ and $\mathfrak G_p(1/q)$. The transposed
orientation enters with relative weight $\kappa=q^{-(L-L_A)}$,
yielding $\mathcal N_R=\mathcal F_R+\kappa\,\mathcal F_{\widetilde R}$.}
\end{figure}

Finally, combining the contributions of the two orientations gives
\begin{equation}
\;\mathcal N_R=\mathcal F_R+\kappa\,\mathcal F_{\widetilde R}\;.
\label{eq:SM_split}
\end{equation}
This is the only conversion formula required in
Eq.~\eqref{eq:resourceconversion}. For dipole-resolved quantities, the actual site positions must be
retained. The transposed dictionary is defined positionwise by
$c_{\widetilde R}(E)=c_R(A\setminus E)$, where $E$ is the set of identity
sites. 
The four steps, together with the structure of
the boundary sensitivity $\mathcal N_R$, are illustrated schematically
in Fig.~\ref{fig:SM_steps}.

\end{document}